\documentclass[usenatbib,useAMS]{mn2e}
\usepackage{natbib}
\usepackage{amsmath}
\usepackage{graphicx}
\usepackage{color}
\usepackage{epsfig}
\usepackage{mathrsfs}
\usepackage{ulem}
\usepackage{times}
\usepackage{balance}
\usepackage{amssymb,amsmath}
\usepackage{longtable,lscape}
\usepackage{url}
\usepackage{bm}
\usepackage{rotating}
\usepackage{hyperref}

\DeclareMathAlphabet{\mathbi}{OT1}{ptm}{bx}{it}
\SetMathAlphabet\mathbi{bold}{OT1}{ptm}{bx}{it}
\def\sss{\scriptscriptstyle}
\def\bhm{M_{\bullet}}

\def\dotM{\dot{M}_{\bullet}}

\def\ergs{\rm erg~s^{-1}}

\def\fblr{f_{\sss\rm BLR}}

\def\kms{\rm km~s^{-1}}

\def\msun{M_{\odot}}

\def\mathdotM{\dot{\mathscr{M}}}
\def\mathD{\mathscr{D}_{\bullet}}

\def\rblr{R_{\sss{\rm BLR}}}
\def\sunm{M_\odot}

\def\mgii{Mg~{\sc ii}}

\def\RSG{R_{_{\rm SG}}}
\def\Rg{R_{\rm g}}
\def\Tobs{T_{\rm obs}}
\def\T5100{T_{\rm 5100}}
\def\TSG{T_{\rm SG}}

\def\TWD{T_{\rm WD}}

\def\mnras{MNRAS}
\def\apj{ApJ}

\def\aap{A\&A}
\def\apjl{ApJ Letters}
\def\apjs{ApJS}

\def\araa{ARA\&A}
\def\nat{Nature}

\begin{document}

\voffset=-0.3in

\title[{Periodicity of Long-term Variations of AGNs}]
{A note on periodicity of long-term variations of optical continuum in
active galactic nuclei}

\author[Lu et al.]
{Kai-Xing Lu$^{1,2}$\thanks{E-mail: lukx@mail.bnu.edu.cn},
Yan-Rong Li$^{2}$,
Shao-Lan Bi$^{1}$ and
Jian-Min Wang$^{2,3,*}$\thanks{E-mail: wangjm@mail.ihep.ac.cn}
\\
$^{1}$Department of Astronomy, Beijing Normal University, Beijing 100875, China\\
$^{2}$Key Laboratory for Particle Astrophysics, Institute of High Energy Physics (IHEP),
Chinese Academy of Sciences, 19B Yuquan Road, Beijing 100049, China\\
$^{3}$National Astronomical Observatories of China, Chinese Academy of Sciences, 20A Datun Road,
Beijing 100020, China\\
}

\date{Accepted 2016 April 01. Received 2016 April 01; in original form 2015 November 15}

\pagerange{\pageref{firstpage}--\pageref{lastpage}} \pubyear{2015} 
\maketitle
\label{firstpage}

\begin{abstract}
Graham et al. found a sample of active galactic nuclei (AGNs) and quasars from 
the Catalina Real-time Transient Survey (CRTS) that have long-term periodic variations 
in optical continuum, the nature of the periodicity remains uncertain. 
 We investigate the periodic variability characteristics of the sample by testing 
the relations of the observed variability periods with AGN optical luminosity, 
black hole mass and accretion rates, and find no significant correlations. 
We also test the observed periods in several different aspects related to accretion 
disks surrounding single black holes, such as the Keplerian rotational periods of 
5100~\AA\, photon-emission regions and self-gravity dominated regions and the precessing period of 
warped disks. These tests shed new lights on understanding AGN variability in general. 
Under the assumption that the periodic behavior is associated with SMBHB systems in particular, 
we compare the separations ($\mathD$) against characteristic radii of broad-line regions ($\rblr$) 
of the binaries and find $\mathD\approx 0.05\rblr$. This interestingly implies that 
these binaries have only circumbinary BLRs. 
\end{abstract}


\section{Introduction}
The optical and ultraviolet spectra of AGNs and quasars have been understood profoundly since their 
discovery. Accretion onto supermassive black holes (SMBHs) located in galactic centres is powering the 
giant emissions of AGNs and quasars through release of gravitational energy of the infalling gas 
\citep{Rees1984}. It is well understood that the broad emission lines arise from gas photoionized by 
accretion disks of SMBHs (\citealt{Osterbrock1986}). According to the hypothesis 
of supermassive black hole binaries (SMBHBs; \citealt{Begelman1980}), there should be some 
AGNs and quasars powered by them at least (\citealt{Gaskell1983}), manifesting with double-peaked or asymmetric 
emission lines (\citealt{Shen2010}) and long-term periodic variations 
(\citealt{Runnoe2015}). Although there are indeed growing indirect evidence for SMBHBs (e.g., \citealt{Yan2015} 
based on special characteristic of spectral energy distributions, or \citealt{Liu2014} on features of 
tidal disruption event in galactic centres), identification of them, in particular in sub-parsec 
scale, is still challenging.

Since long-term monitoring campaigns are extremely time-consuming, 
only a few AGNs and quasars, such as OJ 287 (\citealt{Valtonen2008}), PG 1302-102 (\citealt{Graham2015a}) 
and PSO J334.2028+01.4075 \citep{Liu2015}, have been found to exhibit long-term periodic 
(or quasi) variations of a few to $\sim10$ years in optical and ultraviolet 
continuum. A major advance was made recently by \citet[hereafter G15b]{Graham2015b}, 
who performed a systematic search for the long-term periodical variations of continuum in quasars 
covered by the CRTS and finally identified more than one hundred of candidates. 
Periodic variability is generally believed to be a common signature of SMBHBs, 
but there also exist alternative explanations (\citealt{Graham2015a, Li2016}; G15b). 
The CRTS sample offers an opportunity to test the periodicity and probe the properties of SMBHBs. 
In this paper, we extend the G15b study to show more statistics of the long-term periodicity. 
Throughout this work, we assume a standard $\Lambda$CDM cosmology with $H_0=67~{\rm km~s^{-1}~Mpc^{-1}}$, 
$\Omega_{\Lambda}=0.68$ and $\Omega_{\rm M}=0.32$ (\citealt{Ade2014}). 

\begin{figure*}
\centering
\includegraphics[angle=0,width=0.7\textwidth]{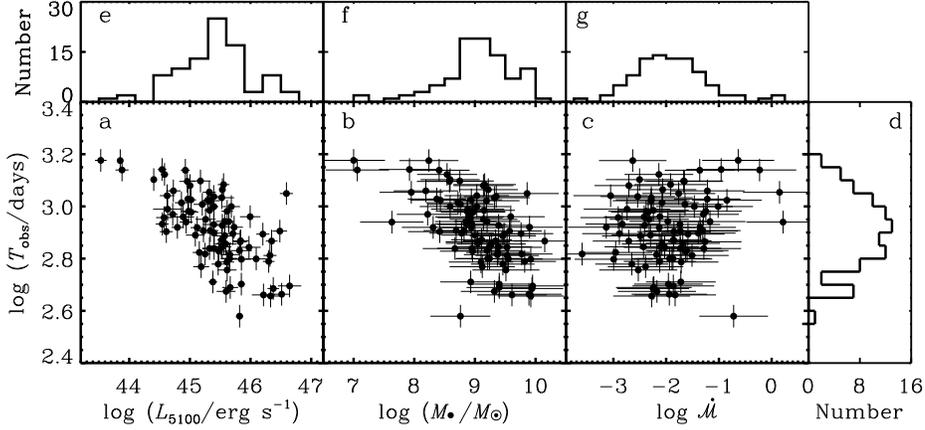}
\caption{
The observed $\Tobs$ in rest-frame versus luminosity ($L_{5100}$), black hole mass ($\bhm$) and 
accretion rates ($\mathdotM$) in panels ({\it a, b, c}). The apparent correlations 
are spurious and caused by the redshift effect 
(see Section 3.1 for details), panels ({\it d, e, f, g}) is the distribution of 
($\Tobs$, $L_{5100}$, $\bhm$, $\mathdotM$)
}
\label{fig:pad}
\end{figure*}

\section{The sample}
The CRTS sample consists of 111 quasars with unambiguous observed 
periods from a few to $\sim 5$ years, as listed in Table~\ref{tab:1} of G15b. 
We cross-checked the sample with Sloan Digital Sky Survey (SDSS) data and 
found available spectra for 90 quasars. Data Reduction of the 
SDSS spectra followed the procedures described in \citet{Hu2008}. 
We measured the 5100~\AA\, luminosity and 
full width at half maximum (FWHM) of the broad H$\beta$ and \mgii\, lines. We used a relation of 
$L_{5100}=0.56L_{3000}$ to convert the ultraviolet continuum into 5100~\AA\, (\citealt{Shen2008}). 

We used the well-established relation between BLR size and AGN 5100~\AA\, luminosity 
($R-L$ relation; \citealt{Bentz2013}) to estimate the BLR size 
\begin{equation}
\rblr\approx 36.3~\ell_{44}^{0.54}~{\rm ltd},
\end{equation}
where $\ell_{44}=L_{5100}/10^{44}~\ergs$ is the luminosity at 5100\AA. We followed the 
standard way of estimating BH mass, 
\begin{equation}
\bhm=\fblr\frac{V_{\rm FWHM}^2\rblr}{G}=5.27\times 10^7~\fblr V_{3000}^2R_{30}\,\sunm,
\end{equation}
where $G$ is the gravitational constant, $R_{30}=\rblr/30$~lt-d, 
$V_{3000}=V_{\rm FWHM}/3000~\kms$ is the FWHM velocity, and $\fblr$ is a 
constant which includes all the unknown information about the geometry and kinematics 
of the BLR. The factor $\fblr$ is obtained by calibrating Equation~(2) against SMBH mass 
obtained from the well known $\bhm-\sigma_*$ relationship in local bulge galaxies. 
We took $\fblr=1.0$ in this paper (\citealt{Ho2014}). 

Note that Equation~(2) only applies to local quasars with H$\beta$ emission line ($z<0.7$). 
For high$-z$ quasars, we used the extended relation of Equation (2) using \mgii\ emission line 
(\citealt{Vestergaard2006}). We then use the obtained  black hole mass 
to calculate the parameters of accretion disks and extensively explore if the observed long-term 
periods are related to accretion disks (see \S3 below). 

\begin{figure*}
\centering
\includegraphics[angle=0,width=0.95\textwidth]{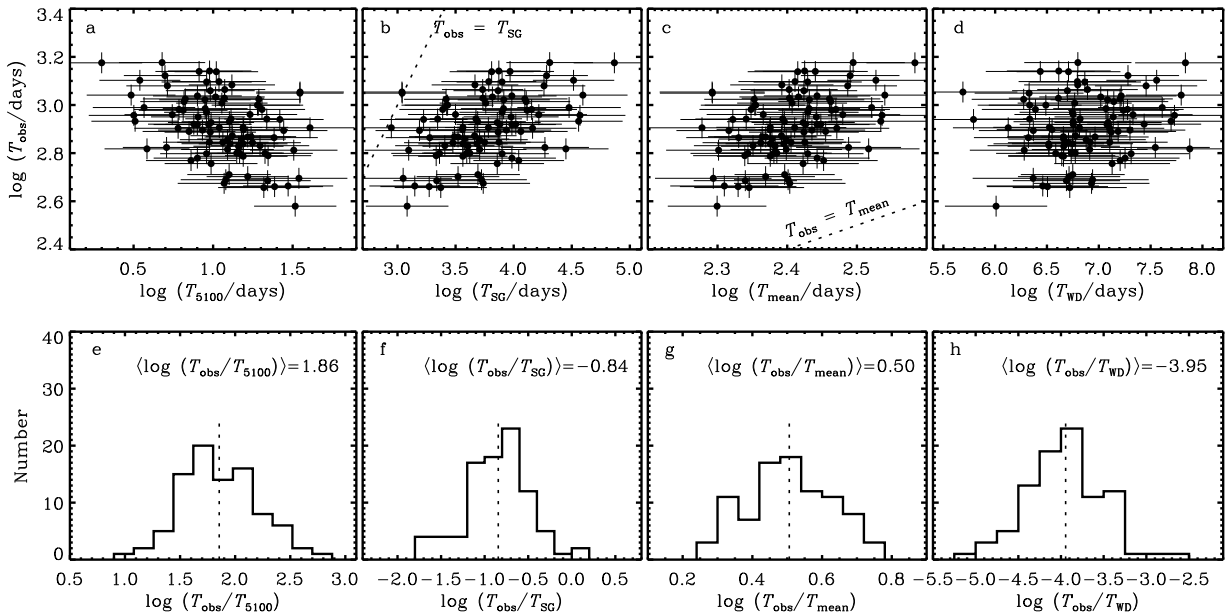}
\caption{
Comparison of observed periods in rest-frame versus the characteristic periods of accretion disks. 
Panels ({\it a, b, c, d}) is the Keplerian periods at the region for 5100\AA\, photons ($\T5100$), 
the critical radius of self-gravity ($\TSG$), averaged radius ($T_{\rm mean}$) and 
the precessing period of warped disks ($\TWD$) versus observed periods($\Tobs$), respectively. 
Their corresponding histograms show in panels ({\it e, f, g, h}). 
Vertical dash lines are the average levels. 
}
\label{fig:pt}
\end{figure*}

\begin{table*}
\centering
\caption{The sample properties. Columns are the objects name, 
redshifts, optical luminosity,
black hole mass, accretion rates, the radii of emitting 5100\AA\, photons and BLR, 
and separations of the black hole binaries calculated from Kepler's law, respectively.
(The full version is available in a machine-readable form in the online.
A portion is shown here for guidance regarding its form and content.)
}
\label{candidates}
\begin{tabular}{lccccccccrccccccc}
\hline
Object & $z$ & log $L_{\rm 5100}$  &  log ($\bhm/\msun$) & log $\mathdotM$ & log ($R_{\rm 5100}/$cm) & log ($R_{\rm BLR}/$cm) & log ($\mathD/$cm) \\
       &     & (erg s$^{-1}$)     &                     &                &               &              &      \\
\hline
                      UM 234  & 0.729 &$ 45.35 \pm  0.12 $&$  9.19 \pm 0.55 $&$  -2.28 \pm   0.73 $&$ 15.16 \pm  0.54 $&$ 17.70 \pm  0.21 $&$ 16.54 \pm  0.18$\\
    SDSS J014350.13+141453.0  & 1.438 &$ 45.65 \pm  0.10 $&$  9.21 \pm 0.53 $&$  -1.87 \pm   0.71 $&$ 15.31 \pm  0.53 $&$ 17.87 \pm  0.20 $&$ 16.40 \pm  0.17$\\
                     US 3204  & 0.954 &$ 45.39 \pm  0.15 $&$  8.95 \pm 0.57 $&$  -1.74 \pm   0.76 $&$ 15.19 \pm  0.57 $&$ 17.73 \pm  0.24 $&$ 16.40 \pm  0.19$\\
    SDSS J072908.71+400836.6  & 0.074 &$ 43.53 \pm  0.08 $&$  7.00 \pm 0.52 $&$  -0.63 \pm   0.68 $&$ 14.25 \pm  0.51 $&$ 16.72 \pm  0.18 $&$ 15.92 \pm  0.17$\\
    \hline
\end{tabular}
\label{tab:1}
\end{table*}

In Table~\ref{tab:1}, we list the main properties of the sample. 
G15b determined the observed periods of the sample by the WWZ$-$ACF method, 
which gives a typical uncertainty of 10 percent for the periods. 
For the uncertainties of black hole mass estimate, 
we included the intrinsic scatters of 0.4~dex (\citealt{Vestergaard2006}). 
For the uncertainties of 5100~\AA\, luminosity $L_{5100}$, we digitized 
the light curves from G15b (see their Figure 7) and included the standard deviation of light curves. 

Figures~\ref{fig:pad}({\it e, f, g, h}) plot the distributions of the observed periods, 
5100~\AA\, luminosities, black hole mass and accretion rates (see \S3 for a definition). 
The current sample shows that 1) the period peaks at $\sim$800 days; 
2) the luminosity cover from $\sim10^{44}~\ergs$ to $\sim10^{47}~\ergs$; 3) the black hole mass at $\sim10^9\sunm$, 
but extends to $10^{10}\sunm$; and 4) the accretion rate spans from $\sim 10^{-3}$ to $10^{-1}$. 
The moderate accretion rates indicate that the objects of the sample are in the regime 
of standard accretion disk model, which is described by \citet{Shakura1973}. 

\section{Statistics of the periods}
As for the accretion disks, we study four characteristic radii and explore 
if they can account for the long-term periodicity. In the standard accretion disk model, 
the effective temperature as a function of disk radius is given by 
\begin{equation}
T_{\rm eff}(R)=\left(\frac{3G\bhm \dotM}{8\pi R^3\sigma_{_{\rm SB}}}\right)^{1/4}
\approx 3.32\times 10^3~\mathdotM^{1/4}M_8^{-1/4}r_3^{-3/4}~{\rm K},
\end{equation}
where $\sigma_{_{\rm SB}}$ is the Stefan-Boltzman constant, $r_3=R/10^3\Rg$, 
$\Rg=1.5\times 10^{13}M_8$~cm is the gravitational radius, 
$\mathdotM=\dotM/L_{\rm Edd}c^{-2}$ is the dimensionless accretion rate, 
and $L_{\rm Edd}=1.26\times 10^{38}\left(\bhm/\sunm\right)\ergs$ is the Eddington luminosity. 
Here we neglect the inner boundary condition. 
The emergent spectra of accretion disks are given by 
$L_{\lambda}\propto \int_{R_{\rm in}}^{\infty} B_{\lambda}(T_{\rm eff})RdR$, where 
$R_{\rm in}$ is the inner radius of the disk and $B_{\lambda}(T_{\rm eff})$ is the Planck function. 
This yields the well-known canonic spectra as $L_{\lambda}\propto \lambda^{-1/3}$. 
It is easy to show that the accretion rate can be expressed by 5100~\AA\, luminosity and black hole mass \citep{Du2014}
\begin{equation}
\mathdotM=0.201\left(\frac{\ell_{44}}{\cos i}\right)^{3/2}M_8^{-2},
\end{equation}
where $\cos i$ is the cosine of the inclination angle of the disk. 
We take $\cos i=0.75$, which corresponds to the opening angle of the dusty torus. 

\subsection{Correlations with $L_{5100}, \bhm$ and $\mathdotM$}
Figure~\ref{fig:pad} show the relation between the rest-frame periods 
with the 5100~\AA\, luminosities, black hole mass and accretion rates, respectively; 
apparent correlations can be seen. However, we note that the sample selection is based on 
both the magnitude limit and the time span of the monitoring ($\sim9$ years). 
Since the redshift distribution of the sample spans a wide range of $z\sim0-3$, 
the relations of the rest-frame periods with AGN parameters will be significantly distorted 
by the redshift effect. Specifically, high-$z$ quasars have higher luminosities but smaller 
rest-frame periods, leading to spurious correlations. We employ partial correlation analysis 
(see the Appendix) to quantitatively test if the correlations are caused by the redshift effect. 
The Spearman rank-order coefficients of the correlation between $\log \Tobs$ 
and ($\log L_{5100}$, $\log \bhm$, $\log \mathdotM$) are $\mathbi{r}=(-0.64, -0.60, 0.14)$, respectively. 
The coefficients of the correlation between ($\log \Tobs$, $\log L_{5100}$, $\log \bhm$, $\log \mathdotM$) and redshift 
are $\mathbi{r}=(-0.73, 0.81, 0.77, -0.16)$, respectively. The partial correlation coefficients between $\log\Tobs$ 
and ($\log L_{5100}$, $\log\bhm$, $\log\mathdotM$) are  $\mathbi{r}=(-0.10, -0.09, 0.03)$ with null probabilities 
of $p=(0.08, 0.10, 0.20)$, respectively. The low values of the partial correlation coefficients confirm 
that the apparent correlations are spurious. There are no significant correlation between the rest-frame period 
and optical luminosity, black hole mass and accretion rate. 

If the periodicity arises from accretion disks of single black holes, 
the observed variability period and our correlation analysis provide useful restraints on 
variablitiy models in accretion disks. For example, \citet{Clarke1989} propsoed a theoretical model 
based on disk thermal instability and predicted that the continuum variations of AGN obey 
a relation of $P\propto L^{1/2}$, and the timescale $>100$ years for the high-luminosity AGNs, 
which is much longer than the observed periods of G15b sample. 

On the other hand, under the SMBHB hypothesis, one may expect that the orbital period would depend on the 
total black hole mass of the system. However, partial correlation analysis shown that there is no correlation 
between the period ($\Tobs$ in rest-frame) and measured black hole mass. The intrinsic scatter of 
black hole mass estimated from the $R-L_{5100}$ relation is as large as 0.43~dex (e.g., \citealt{Vestergaard2006}). 
Meanwhile, the possible biases between measured black hole mass and ``true'' mass may lead to additional scatters 
(e.g., \citealt{Shen2008}). It is plausible that the intrinsic correlation between the periods 
and black hole mass is smeared out due to the large scatter of mass measurements.

\subsection{Characteristic periods of accretion disks surrounding single black holes}
We now turn to calculate the following four characteristic periods of accretion disks 
surrounding single black holes to test if they can account for the observed periods.

For the standard accretion disk model, the photons with a wavelength $\lambda$ mainly come from 
the region with a temperature of $T_{\rm eff}=2.37\times 10^8/\lambda({\rm \AA})$~K 
(\citealt{Siemiginowska1989}). We thus have the region for 5100~{\AA} photons 
$R_{5100}/\Rg=29.7\mathdotM^{1/3}M_8^{-1/3}\lambda_{5100}^{4/3}$ from Equation~(4), namely, 
\begin{equation}
\frac{R_{5100}}{\Rg}=17.4\left(\frac{\ell_{44}}{\cos i}\right)^{1/2}M_8^{-1}\lambda_{5100}^{4/3},
\end{equation}
where $\lambda_{5100}=\lambda/5100{\rm \AA}$. 

The outer part of accretion disks are optically thick and geometrically thin, dominated by gas 
pressure. In a pioneering paper, \citet{Paczynski1978} realized that this region is so distant that it is 
strongly affected by the vertical self-gravity of the disk rather than the central black hole. The self-gravity 
of the disk is $2\pi G\Sigma$, while the vertical force of the gravity is given by $G\bhm R^{-2}(H/R)$. 
Using the mass density of the outer part of accretion disks (the region~C solution in \citealt{Shakura1973}) 
\begin{equation}
\rho=4.38\times10^{-9}~\alpha_{0.1}^{-7/10}\mathdotM^{11/20}M_8^{-7/10}r_3^{-15/8} {\rm g~cm^{-3}},
\end{equation}
we have the Toomre parameter, defined by the ratio of the vertical force of the central black hole 
to the disk's self-gravity, as 
\begin{equation}
Q=\frac{\bhm}{4\pi \rho R^3}=1.10~\alpha_{0.1}^{7/10}M_8^{-13/10}\mathdotM^{-11/20}r_3^{-9/8},
\end{equation}
where $\alpha_{0.1}=\alpha/0.1$. When $Q\lesssim 1$, 
the disks become self-gravity dominated. This yields a critical radius of self-gravity as 
\begin{equation}
\frac{\RSG}{R_{\rm g}}=1.09\times 10^3~\alpha_{0.1}^{28/45}M_8^{-52/45}\mathdotM^{-22/45}.
\end{equation}
With two radii of Equation (5) and (8), the corresponding Keplerian periods are given by 
\begin{equation}
T_{5100, {\rm SG}}=2\pi\sqrt{\frac{(R_{5100},R_{\rm SG})^3}{G\bhm}},
\end{equation}
where the subscripts refer to the corresponding radii. 
In addition, to account for the region between $R_{5100}$ and $\RSG$, we define an averaged 
radius as $\langle R\rangle=\sqrt{R_{5100}\RSG}$ and the averaged period as 
\begin{equation}
\langle T\rangle=2\pi \sqrt{\frac{\langle R\rangle^3}{G\bhm}}.
\end{equation}

At last, we can also calculate the characteristic period for warped disks. 
According to \cite{Shakura1973}, the total mass of the disk within $\RSG$ is 
\begin{equation}
M_{\rm disk}\approx 2\pi \RSG^2\Sigma_{_{\rm SG}}\approx 1.42\times 10^5~\alpha_{0.1}^{-1/45}\ell_{44}^{6/45}M_8^{26/45}~\sunm,
\end{equation}
where $\Sigma_{_{\rm SG}}$ is the surface density of the disk at $\RSG$. 
The warped disks precess with a typical period 
\begin{equation}
T_{\rm WD}=2\pi \sqrt{\frac{\bhm R_{\rm WD}^3}{GM_{\rm disk}^2}},
\end{equation}
where $R_{\rm WD}$ is typical radius ($R_{\rm WD}$=$\RSG$) of the warped disk (\citealt{Ulubay-Siddiki2009}). 

Figure~\ref{fig:pt} compares the rest-frame $T_{\rm obs}$ with $T_{5100}$, $T_{\rm SG}$, 
$T_{\rm mean}$ and $T_{\rm WD}$ in the top panels, and plots their respective ratios in the bottom panels. 
We find that $T_{\rm obs}$ is much longer than $T_{5100}$, but shorter than $T_{\rm SG}$, 
namely $T_{\rm 5100}<\Tobs<T_{\rm SG}$. This motivates us to compare $T_{\rm obs}$ with the mean value of rotational periods 
between $R_{\rm 5100}$ and $R_{\rm SG}$. As shown by Figure~\ref{fig:pt}{\it g}, $T_{\rm obs}$ is 
greater than $T_{\rm mean}$, implying that $T_{\rm obs}$ is related to the region between $R_{\rm SG}$ and 
$\langle R\rangle$ if the observed periodicity arises from an accretion disk surrounding a single black hole. 
Figure~\ref{fig:pt}{\it h} shows that $\Tobs$ is much shorter than the period of warped disks driven by the 
self-gravity torque, so that we can rule out this possibility for the periodicity. 

In next section, under the assumption that the observed periods are associated with SMBHB systems, 
we discuss several possible processes for the periodicity and 
calculate the BLR sizes and separations of these SMBHB systems.

\section{SMBHB models}
\subsection{Periodicity explanations}
As pointed out by G15b, there exist several plausible processes responsible for the periodicity. 
\begin{enumerate}
 \item {\it A processing jet}. In the sample, by checking the radio band data from 
 the catalog of VLA-FIRST survey \citep{Chang2004}, we found that only 13 candidates are radio-loud, 
 indicating that the periodic variations in most candidates are not caused by a precessing jet. 
 
 \item {\it Quasi-periodic oscillation}. \citet{Abramowicz2001} suggested that high-frequency QPOs 
 arise from some type of resonance mechanism. 
 \citet{Zhou2010} showed that the observed time-scales ($T$) of high-frequency QPOs 
 correlate with black hole mass following $T=6.22~\bhm/~(10^{9}~\sunm)$~days. 
 For G15b sample, the typical value ($T\sim6.22$ days) is much shorter than observed period. 
 On the other hand, as for low-frequency QPOs, \citet{Yan2013} showed 
 that GRS 1915+105 has a mass of $\sim12\sunm$ and exhibits QPO with 1 Hz. 
If we assume that the period $T\propto M_\bullet$,  the case of GRS 1915+105 would mean that 
the period if QPOs of G15b sample, spanning a mass range of $10^8-10^{10}M_\odot$, lies in a range 
between $\sim100$ to $10^{4}$ days (in observed-frame). This is generally comparable with the range of 
observed periods in G15b sample. 

 \item {\it Periodic accretion.} The orbital motion of binaries induces periodic 
 modulation of mass accretion onto each black hole \citep{Farris2014}, which translates into periodic 
 emissions of the accretion disks. The issue about this explanation is that the viscous time of 
 an accretion disk, which reflects the timescale of its response to a change in mass accretion rate, 
 is generally much longer than the observed periods of the present sample. 
 
 \item {\it Relativistic Doppler boosting}. Recently, \cite{D'Orazio2015} proposed an alternative 
 explanation for the periodicity by relativistic Doppler boosting and applied it to the periodic 
 light curve of PG~1302-102. In this model, a large inclination of the binary's orbit is generally required 
 to account for the variation amplitude. This will rise a concern about the obscuration of the outer 
 dusty torus if the dusty torus is coplanar with the binary's orbit. 
  
 \item {\it Warped accretion disk}. The accretion disks are warped due to tidal torques if their 
 spin axis is misaligned with the orbital axis of the binaries. The warps precess along with  
 the binary's orbital motion and eclipse some parts of the disk emissions, leading to periodic variations in 
 the disk emissions. 
\end{enumerate}
In a nutshell, we can only generally exclude with certainty the possibility of the precessing jet model 
and high-frequency QPOs for the periodicity. For the other models, we need more observations, 
in particular spectroscopic data, to test them. 

\begin{figure}
\centering
\includegraphics[angle=0,width=0.47\textwidth]{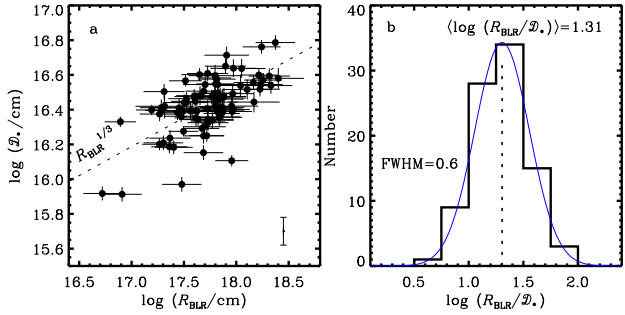}
\caption{
Comparison of SMBHB separations ($\mathD$) with the sizes of broad-line regions ($\rblr$) in panels ({\it a, b}). 
In panel ({\it a}), dotted line is the expected relation. The bar plotted in the right lower corner 
represents the typical scatters of $\mathD$. 
In panel ({\it b}), blue line is the Gaussian fit to the distribution.
}
\label{fig:rr}
\end{figure}

\subsection{Separations and BLR sizes}
Under the assumption that the periodicity arises from a SMBHB system, we now calculate the binary's 
separation, given with the observed period, and compare it against the BLR size predicted from the $R-L$ relation. 
Using the Kepler's law of $\mathD=\left(G\bhm T_{\rm obs}^2/4\pi^2\right)^{1/3}$ 
and Equation (2), we have the binary's semi-major axis as 
\begin{equation}
\mathD=\left(\frac{\fblr}{4\pi^2}\right)^{1/3}\left(V_{\rm FWHM}^2T_{\rm obs}^{2}\rblr\right)^{1/3}.
\end{equation}
This indicates $\mathD\propto \rblr^{1/3}$ for given $V_{\rm FWHM}$ and $\Tobs$ (Figure~\ref{fig:rr}{\it a}). 
We can also rewrite Equation (13) into 
\begin{equation}
\frac{\mathD}{\rblr}
=0.07~\fblr^{1/3}V_{3000}^{2/3}T_{1}^{2/3}R_{30}^{-2/3},
\end{equation}
where $T_{1}=T_{\rm obs}/{\rm year}$. 
Figure~\ref{fig:rr}{\it b} shows that $\mathD$ is smaller than $\rblr$ and in average, $\mathD\approx 0.05 \rblr$ 
in the sample. The FWHM of the $\log \left(\rblr/\mathD\right)$ distribution is 0.6. 
This interestingly means that the BLRs of the binary have been merged, 
but the binary black holes are still co-rotating. 

\section{Conclusions}
We have tested the periodicity of long-term optical variations in a sample of AGNs from G15b. 
The partial correlation analysis shows that the observed periods of G15b sample are uncorrelated with 
the AGN 5100~\AA\, luminosity, (total) black hole mass and accretion rates. 
By comparing the observed periods with the characteristic periods of accretion disks 
surrounding single black holes, we found that the periods generally lies within 
the Keplerian periods that correspond to the regions between the 5100~\AA\, region and the self-gravity radius. 
We discussed several existing explanations for the periodicity in the context of SMBHB hypothesis 
and concluded that further observations (particularly long-term spectroscopic monitoring) are required to test 
if the periodicity originate from SMBHB systems (e.g., \citealt{Li2016}). 
Nevertheless, by assuming that observed periodicity is associated with SMBHB systems, 
we calculated the SMBHB's separations and found that they are smaller than the sizes of broad-line regions, 
implying that the binary BLRs have been merged. 

\section*{Acknowledgements}
We are grateful to the referee for constructive suggestions that significantly 
improved the manuscript. We thank the IHEP AGN members for useful discussions. 
This research is supported by NSFC-11173023, -11133006, -11233003, -11273007 and -11573026. 

\appendix
\section{Partial correlation analysis}
For the parameter sets ($\mathbi{x_{i}, y_{i}, z_{i}}$), $i=1,2...,N$. The correlation coefficient 
between $\mathbi{x}$ and $\mathbi{y}$ excluding the dependence on the third parameter of $\mathbi{z}$ 
is evaluated as (e.g., \citealt{Kendall1979}) 
\begin{equation}
 \mathbi{r_{xy,z}}=\frac{\mathbi{r_{xy}}-\mathbi{r_{xz}r_{yz}}}{\sqrt{1-\mathbi{r_{xz}}^{2}}\sqrt{1-\mathbi{r_{yz}}^2}}, 
\end{equation}
where $\mathbi{r_{xy},~r_{xz}~{\rm and}~r_{yz}}$ is the Spearman rank-order correlation coefficient 
between $\mathbi{x}$ and $\mathbi{y}$, between $\mathbi{x}$ and $\mathbi{z}$ and between $\mathbi{y}$ 
and $\mathbi{z}$, respectively \citep{Press1992}. $\mathbi{r_{xy,z}}$ is partial correlation coefficients.

\end{document}